\documentclass[journal]{IEEEtran}

\IEEEoverridecommandlockouts 

\usepackage{amsmath,amsfonts,amssymb,graphicx,lscape,indentfirst,
latexsym,multirow,nicefrac}
\newtheorem{thm}{Theorem}
\newtheorem{cor}{Corollary}
\newtheorem{lem}{Lemma}
\newtheorem{prop}{Proposition}
\newtheorem{defn}{Definition}
\newtheorem{expl}{Example}

\newcommand\ff{{\mathbb F}}
\def\dist{\qopname\relax{no}{dist}}
\newcommand{\remove}[1]{}
\newcommand\nc\newcommand
\nc\bfa{{\boldsymbol a}}\nc\bfA{{\boldsymbol A}}\nc\cA{{\mathcal A}}
\nc\bfb{{\boldsymbol b}}\nc\bfB{{\boldsymbol B}}\nc\cB{{\mathcal B}}
\nc\bfc{{\boldsymbol c}}\nc\bfC{{\boldsymbol C}}\nc\cC{{\mathcal C}}
\nc\bfd{{\boldsymbol d}}\nc\bfD{{\boldsymbol D}}\nc\cD{{\mathcal D}}
\nc\bfe{{\boldsymbol e}}\nc\bfE{{\boldsymbol E}}\nc\cE{{\mathcal E}}
\nc\bff{{\boldsymbol f}}\nc\bfF{{\boldsymbol F}}\nc\cF{{\mathcal F}}
\nc\bfg{{\boldsymbol g}}\nc\bfG{{\boldsymbol G}}\nc\cG{{\mathcal G}}
\nc\bfh{{\boldsymbol h}}\nc\bfH{{\boldsymbol H}}\nc\cH{{\mathcal H}}
\nc\bfi{{\boldsymbol i}}\nc\bfI{{\boldsymbol I}}\nc\cI{{\mathcal I}}
\nc\bfj{{\boldsymbol j}}\nc\bfJ{{\boldsymbol J}}\nc\cJ{{\mathcal J}}
\nc\bfk{{\boldsymbol k}}\nc\bfK{{\boldsymbol K}}\nc\cK{{\mathcal K}}
\nc\bfl{{\boldsymbol l}}\nc\bfL{{\boldsymbol L}}\nc\cL{{\mathcal L}}
\nc\bfm{{\boldsymbol m}}\nc\bfM{{\boldsymbol M}}\nc\cM{{\mathcal M}}
\nc\bfn{{\boldsymbol n}}\nc\bfN{{\boldsymbol N}}\nc\cN{{\mathcal N}}
\nc\bfo{{\boldsymbol o}}\nc\bfO{{\boldsymbol O}}\nc\cO{{\mathcal O}}
\nc\bfp{{\boldsymbol p}}\nc\bfP{{\boldsymbol P}}\nc\cP{{\mathcal P}}
\nc\bfq{{\boldsymbol q}}\nc\bfQ{{\boldsymbol Q}}\nc\cQ{{\mathcal Q}}
\nc\bfr{{\boldsymbol r}}\nc\bfR{{\boldsymbol R}}\nc\cR{{\mathcal R}}
\nc\bfs{{\boldsymbol s}}\nc\bfS{{\boldsymbol S}}\nc\cS{{\mathcal S}}
\nc\bft{{\boldsymbol t}}\nc\bfT{{\boldsymbol T}}\nc\cT{{\mathcal T}}
\nc\bfu{{\boldsymbol u}}\nc\bfU{{\boldsymbol U}}\nc\cU{{\mathcal U}}
\nc\bfv{{\boldsymbol v}}\nc\bfV{{\boldsymbol V}}\nc\cV{{\mathcal V}}
\nc\bfw{{\boldsymbol w}}\nc\bfW{{\boldsymbol W}}\nc\cW{{\mathcal W}}
\nc\bfx{{\boldsymbol x}}\nc\bfX{{\boldsymbol X}}\nc\cX{{\mathcal X}}
\nc\bfy{{\boldsymbol y}}\nc\bfY{{\boldsymbol Y}}\nc\cY{{\mathcal Y}}
\nc\bfz{{\boldsymbol z}}\nc\bfZ{{\boldsymbol Z}}\nc\cZ{{\mathcal Z}}
\nc\bfal{{\boldsymbol \alpha}}
\renewcommand {\QED} {\QEDopen}

\title{\LARGE \bf
Performance Analysis of Algebraic Soft-Decision Decoding of
Reed-Solomon Codes\thanks{Presented in part at the 44th Annual
Allerton Conference on Communication, Control and Computing,
Monticello, IL, Septermber 27-29, 2006.}}
\author{Andrew~Duggan$^\ast$ and
        Alexander~Barg$^{\text{\S}}$%
\thanks{$^\ast$ E-mail address: aduggan@alum.mit.edu.}%
\thanks{$^{\text{\S}}$ Dept. of. ECE and Institute for Systems
Research, University of Maryland, College Park, MD 20742,
abarg@umd.edu. Supported in part by
NSF grants CCF0515124, CCF0635271, and by NSA grant H98230-06-1-0044.}%
}

\begin{document}

\maketitle \thispagestyle{empty} \pagestyle{empty}

\begin{abstract}

We investigate the decoding region for Algebraic Soft-Decision
Decoding (ASD) of Reed-Solo\-mon codes in a discrete, memoryless,
additive-noise channel. An expression is derived for the error
correction radius within which the soft-decision decoder produces a
list that contains the transmitted codeword. The error radius for
ASD is shown to be larger than that of Guruswami-Sudan (GS)
hard-decision decoding for a subset of low-rate codes. These results
are also extended to multivariable interpolation in the sense of
Parvaresh and Vardy. An upper bound is then presented for ASD's
probability of error, where an error is defined as the event that
the decoder selects an erroneous codeword from its list. This new
definition gives a more accurate bound on the probability of error
of ASD than the results available in the literature.
\end{abstract}


\section{Introduction}

Reed-Solomon (RS) codes are used in a wide variety of applications
currently, and the classical algorithm of Berlekamp and Massey (BM
algorithm) has been employed for decoding in most cases. For an RS
code of length $n$ and dimension $k$, this algorithm is guaranteed
to recover the transmitted codeword within an error radius of
$\lfloor\frac{n-k+1}{2}\rfloor$.

Guruswami and Sudan \cite{GS} presented an important new algebraic
decoding method for RS codes that is able to correct errors beyond
the BM decoding radius. This method involves constructing a
bivariate polynomial with zeros of multiplicity based on the
received symbols. The polynomial can then be factored to give a list
of candidate codewords; thus, it is a list decoder. A
Guruswami-Sudan (GS) decoder includes the transmitted codeword on
its output list if the errors fall within a radius of $n-\sqrt{nk}$.

In \cite{GS}, the authors mention that their hard-decision algebraic
decoding technique can be extended to soft-decision decoding by
setting the values of the multiplicities based on channel posterior
probabilities and not received symbols. However, \cite{GS} does not
provide a way of assigning these multiplicities, which turns out to
be a nontrivial component in the RS decoding procedures. Koetter and
Vardy refined the Algebraic Soft-Decision Decoding (ASD) approach in
\cite{KV} by providing an algorithm that converts a matrix of
posterior probabilities $\Pi$ to a multiplicity assignment matrix
$\mathcal{M}$.

Various papers, such as \cite{PV}, \cite{EM}, \cite{JN}, have been published
since \cite{KV} that propose using a different method for converting
$\Pi$ to $\mathcal{M}$. However, few papers have given insight into
the decoding region of ASD. One exception is \cite{JN}, which
characterizes the decoding region for medium to high rate codes over
binary erasure and binary symmetric channels. Another paper
\cite{KO} derives an error correction radius for an arbitrary
additive cost function associated with transitions in the channel.

This paper examines the performance of ASD when the noise is
additive, i.e., there is a probability distribution $p(a)$ on $q$-ary
errors that does not depend on the transmitted sequence. Relying on
this distribution, we derive in Sect.~\ref{sect:pfm} 
simple estimates of the error radius
within which the transmitted codeword is guaranteed to be on the
list produced by ASD.  With the simple expression for the error
radius obtained, we are also able to characterize the region where
the ASD algorithm provides an improvement over the GS decoding
radius.

In Section \ref{sect:pe}, we study bounds on the probability of
error for ASD decoding. Prior work \cite{RK} has concentrated on the
list-decoding error event, namely the event that the transmitted
codeword fails to be included in the list. We point out that for low
code rates, the list decoding error criterion does not provide
insight into the performance of the decoder because the transmitted
codeword always will be included in the list. We therefore define
decoding to be successful if the ASD algorithm selects the
transmitted codeword from the decoder's list, and we derive an upper
bound for the probability of error based on this new definition.
Finally, in Section \ref{sect:mv} we briefly discuss soft-decision
decoding of multivariate RS codes introduced in a recent work of
Parvaresh and Vardy \cite{VP} and extend to this case our estimate
of the ASD error radius.

The error bounds obtained in Sections \ref{sect:pfm}-\ref{sect:mv}
are either the first of their kind available in the literature, or,
as in the case of the ASD list-decoding error bound, improve the
results known previously.

\section{Decoding of RS codes}\label{sect:RSM}

\subsection{Notation}
Let $q$ be a prime power, let $\mathbb{F}_q = \{\alpha_1 = 0,
\alpha_2, \dots, \alpha_q \}$ be the finite field of $q$ elements,
and let $n=q-1$ be the code length. Denote $\dist(\cdot,\cdot)$ as
the Hamming distance. With a vector $\bfv=(v_1,\dots,v_n)\in
\ff_q^n$ we associate an integer-valued function $v(i)$ whose value
is $j$ if $v_i=\alpha_j.$

For a polynomial $f\in \mathbb{F}[x]$ define the evaluation mapping
$\textnormal{eval}:f \to \mathbb{F}^n_q$ given by
$(\textnormal{eval} f)_i = f(\alpha_i)$, $2 \le i \le q$. Thus, the
evaluation mapping associates a $q$-ary $n$-vector to every
polynomial $f\in \mathbb{F}_q[x].$

\begin{defn}
A $q$-ary RS code $C$ of length $n=q-1$ and dimension $k$ is the set
of codewords of the form
  $$
     \{\bfc=\textnormal{eval} (f): \;f\in \mathbb{F}_q[x], 0\le \deg
f\le k-1
        \}.
  $$
\end{defn}

\smallskip
To describe the encoding of the code $C$, suppose that the message
to be transmitted is $\bfu = (u_1, u_2, \dots, \linebreak[2]u_k)$
where $u_i \in \mathbb{F}_q$, $1\le i \le k.$ The codeword that
corresponds to it is given by $\bfc= \textnormal{eval}(f),$ where
the polynomial $f$ has the form
\begin{equation*}
f(X) = u_1 + u_2 X + u_3 X^2 + \dots + u_kX^{k-1}.
\end{equation*}

We assume that the codeword $\bfc$ is transmitted over a discrete
memoryless channel. In the hard-decision case, the output of the
channel is the vector $\bfy=\bfc+\bfe.$ Let $w_{i,j}=\Pr(y =
\alpha_i | c = \alpha_j)$ be the probability that the symbol
$\alpha_j$ transmitted over the channel is received as $\alpha_i.$
We will assume that the noise is {\em additive}, i.e. there exists a
probability distribution $p$ on $\mathbb{F}_q$ such that
$\Pr(\alpha+e|\alpha)=p(e), \alpha,e\in \ff_q.$ Note that under
these assumptions, the channel is symmetric as defined in Section
8.2 of \cite{CT}.

In the setting of soft-decision decoding, the demodulator is assumed
to provide the decoder with the posterior probabilities conditioned
on the received (continuous) signal. However, the task of analyzing
this setting in the context of algebraic list decoding so far has
proved elusive. We will therefore assume that the receiver outputs
in each position of the codeword, a hard-decision symbol (for
instance, the most likely transmitted symbol) {\em together} with
the $q$ values of {\em posterior probabilities} $\pi_{i,j}=\Pr(c =
\alpha_i|y = \alpha_j)$. As customary in the literature, we will
assume that ASD takes the channel's output to be in the form of a
$q\times n$ matrix $\Pi=[\pi_{i,j}].$

This paper concentrates on the ASD algorithm of \cite{KV} and
compares its performance to the well-known hard-decision decoding
algorithms of Berlekamp and Massey (see, {\em e.g.}, \cite{RB}) and
Guruswami and Sudan \cite{GS,MC}.

\subsection{Hard-Decision Decoding Methods}

Under BM decoding, if the number of errors $t =
\textnormal{dist}(\bfc,\bfy)$ satisfies
\begin{equation}
\label{eq:a} t \le \left\lfloor \frac{n-k+1}{2} \right\rfloor,
\end{equation}

\noindent then the decoder will output $\bfc$. If condition
(\ref{eq:a}) is not true, then decoding is guaranteed to fail.
Therefore, (\ref{eq:a}) is a necessary condition for BM decoding
success.

GS decoding produces a list that contains all the codewords of a
certain distance $t_{m}$ from the vector $\bfy$ and potentially some
codewords outside of this Hamming ball. List decoding success is
declared if the correct codeword is on the list. The distance
$t_{m}$ is determined by $m$ which is a parameter of the algorithm.
As $m$ increases, $t_m$ increases to an asymptotic limit given in
Lemma \ref{lem:a}.

\begin{lem}
\textnormal{(Guruswami and Sudan \cite{GS})} Let $m \to \infty.$ Let
$\bfc$ be a codeword that satisfies
\begin{equation}\label{eq:b}
\textnormal{dist}(\bfy,\bfc) < n - \sqrt{nk}.
\end{equation}
Then $\bfc$ will be included in the list output by the GS decoder
with input $\bfy.$ \label{lem:a}
\end{lem}

The complexity of the algorithm often becomes a limiting factor
before the maximum possible $t_m$ is achieved. Note that
(\ref{eq:b}) is only a sufficient condition on GS list-decoding
success: the decoding is guaranteed to have the transmitted codeword
on the list if the error pattern satisfies (\ref{eq:b}), assuming
large $m$.

Let $\tau = \nicefrac{t}{n}$ be the normalized error correction
radius of RS decoding algorithms, and let $R = \nicefrac{k}{n}$ be
the rate of the code. We have
\begin{equation}\label{eq:new}
\begin{split}
& \tau = \frac{1-R}{2} \; \; \textnormal{(BM Decoding)} \\
& \tau = 1 - \sqrt{R} \; \; \textnormal{(GS Decoding, } m
\textnormal{ large).}
\end{split}
\end{equation}

\noindent The two error radii given in (\ref{eq:new}) are shown as
the dashed curves in Figure 1(a). The GS decoding radius is always
greater than its BM counterpart although the difference becomes
small for high rates. However, no error radius for Algebraic
Soft-Decision Decoding was previously known.

\subsection{ASD Algorithms}

ASD is an extension of GS decoding by the manipulation of the
multiplicities. We only mention the salient features of this
algorithm, referring to \cite{KV} for the details. Instead of
operating on a vector $\bfy$, ASD takes as input a multiplicity
matrix $\mathcal{M}$ of dimensions $q \times n$ constructed on the
basis of the posterior probability matrix $\Pi.$ The soft-decision
decoder constructs a bivariate polynomial $Q(X,Y)$ that has zeros of
multiplicity set by $\mathcal{M}.$ In the next step, the decoder
recovers a list $\cL$ of putative transmitted codewords from the
$Y$-zeros of $Q(X,Y)$. In contrast to GS decoding that always
constructs $Q(X,Y)$ based on $n$ distinct zeros, ASD can have up to
$qn$ distinct zeros.

\remove{Similarly to GS decoding, ASD uses curve-fitting of codeword
polynomials to points. However, since there are many more points,
including multiple points for the same codeword position, Hamming
distance is an inadequate way to measure the fit of a codeword.
Thus, we must introduce the notion of a codeword score that tracks
the fit of the codeword candidates to the multiplicities in $\cM.$}

\begin{defn}
Define the \textit{Score} of a vector $\bfv=(v_1,\dots,v_n)\in
\ff_q^n$ with respect to a multiplicity matrix $\mathcal{M}$ to be
\begin{equation*}
S_\mathcal{M}(\bfv) = \sum^n_{i=1}m_{v(i),i}.
\end{equation*}
where $v(i)=\ell$ if $v_i=\alpha_\ell.$
\end{defn}

If needed, in the last stage the decoder chooses the codeword $\bfc$
from the list $\cL$ with the largest score or the codeword with the
largest probability $P^n(\bfc|\bfy)=\prod \pi_{c_i,y_i}$.

\subsubsection{The Multiplicity Matrix}

The matrix $\mathcal{M}$ is determined from the matrix of posterior
probabilities $\Pi$. Koetter and Vardy \cite{KV}, Parvaresh and
Vardy \cite{PV}, and El-Khamy and McEliece \cite{EM} have proposed
various methods for determining $\mathcal{M}$ from $\Pi$. A simple
method for converting $\Pi$ to $\mathcal{M}$ that will be used in
this paper is called the Proportionality Multiplicity Assignment
Strategy (PMAS) proposed by Gross et al. \cite{GK}. PMAS finds
$\mathcal{M}$ by performing the following element-wise calculation
on $\Pi$ for some fixed number $\lambda:$
\begin{equation}\label{eq:i}
m_{i,j} = \lfloor\lambda\pi_{i,j}\rfloor, \quad i=1,\dots, 1;
\,j=1,\dots, n.
\end{equation}
Thus, the matrix $\cM$ is determined uniquely from the received
vector $\bfy$ and the properties of the communication channel. The
parameter $\lambda \in \mathbb{Z}^+$ is the complexity factor, and
its adjustment controls directly the balance between the performance
and the complexity of ASD. Another important measure of the
complexity of the decoder is the cost of the multiplicity matrix,
defined as follows.

\begin{defn}
Let the \textit{Cost} of a multiplicity matrix $\mathcal{M}$ be
\begin{equation*}
\mathcal{C}(\mathcal{M}) = \frac{1}{2}\sum_{i,j}m_{i,j}(m_{i,j} +
1).
\end{equation*}
\end{defn}




\subsubsection{Threshold Condition}

GS decoding includes on its output list all codewords in the Hamming
space within a certain radius of the received vector $\bfy$, but ASD
has no known geometric interpretation. A sufficient condition for
ASD's success is determined indirectly by the vector $\bfy$ and can
be stated in terms of the score $S_{\cM}(\bfc)$ as given in the next
lemma.
\begin{lem}\label{lem:h}
\textnormal{\cite{KV}} Suppose ASD is used to decode a received
vector $\bfy$ with the RS code. If
\begin{equation*}
S_{\mathcal{M}}(\bfc) > \sqrt{2(k-1)\mathcal{C}(\mathcal{M})}
\end{equation*}
\noindent or equivalently
\begin{equation*}
\sum^n_{i=1}m_{c(i),i} > \sqrt{(k-1)\sum_{i,j}m_{i,j}(m_{i,j} + 1)},
\end{equation*}
\noindent then the transmitted codeword is on the decoder's list.
\end{lem}

Lemma \ref{lem:h} can be used to evaluate ASD's performance in the
case that $\bfc$ is transmitted, $\bfy$ is received, and
$\mathcal{M}$ is the multiplicity matrix.

\subsubsection{Size of the List}

The decoder's list $\mathcal{L}$ cannot exceed the $Y$-degree of
$\mathcal{Q}(X,Y)$ since $\mathcal{L}$ is obtained by factoring
$\mathcal{Q}(X,Y)$ for the $Y$-roots. The size of the list in terms
of the $Y$-degree $L$ of $\cQ$ is estimated in the next lemma, which
is due to McEliece \cite{MC}, Corollary 5.14.
\begin{lem}\label{lem:him}
For $k \ge 2 $, the number of codewords on the list of an algebraic
soft-decision decoder does not exceed

\begin{equation*}\label{eq:ppo}
L = \left\lfloor\sqrt{\frac{2\mathcal{C}(\mathcal{M})}{k-1} +
\left(\frac{k+1}{2k-2}\right)^2}-\left(\frac{k+1}{2k-2}\right)\right\rfloor.
\end{equation*}
%
%
\end{lem}
%
The condition $| \mathcal{L} | \le L$ is met with equality if and
only if all the $Y$-roots have degree less than $k$.

\section{ASD Error Correction Performance}\label{sect:pfm}

In this section, we present one of our main results, an estimate of
the error correction radius $t$ of the algorithm. We would like to
stress one essential difference of the result below from the other
similar results in the literature.  In the case of BM and GS
decoding for instance, all of the codewords within the error radius
are included in the list output by the decoder. In contrast, we only
guarantee in the case of ASD that if the {\em transmitted} codeword
is distance $t$ away from the received one, then it will be included
on decoder's list. Other codewords, even within the sphere of radius
$t$ from $\bfy$, may escape being output by the decoder. In other
words, the decoding regions of ASD are far from being spherical, and
in fact, no geometric characterization of them is available.

\subsection{Setting}
Following Koetter and Vardy in \cite{KV} and Justesen in \cite{JU},
we assume that each symbol entering the channel is uniformly drawn
from $\mathbb{F}_q$. It follows that $\pi_{i,j} = w_{i,j}$. Since
the channel is symmetric, we know that the channel transition
probabilities $w_{i,j}$ are drawn from the set $\{p_1, p_2, \dots,
p_q\}$. Next, we will introduce the three channel statistics
\begin{equation*}
p_{\max} = \displaystyle\max_{1 \le i \le q} p_i \; \; \; \; \; \;
p_{\min} = \displaystyle\min_{\substack{1 \le i \le q \\
i:p_i > 0}} p_i \; \; \; \; \; \; \gamma = \displaystyle\sum^q_{i=1}
p^2_i.
\end{equation*}
When the channel is noiseless, set $p_{\min} = 0$. We will assume
throughout that the channel's capacity is greater than zero, giving
us $p_{\max} > p_{\min}$. As will be seen later, $p_{\max}$,
$p_{\min}$, and $\gamma$ will be the only channel statistics
necessary in our analysis of ASD's performance.

\subsection{Error Radius}

\begin{thm}\label{thm:a}
Suppose that an RS code with rate $R = \nicefrac{k}{n}$ is used to
communicate over an discrete, additive-noise channel. Suppose that a
codeword $\bfc$ is transmitted and an algebraic soft-decision
decoder with complexity factor $\lambda$ is used to decode the
received vector $\bfy.$ Let $t = \textnormal{dist}(\bfc,\bfy)$. If
\begin{equation}\label{eq:t}
\frac{t}{n} \le \frac{p_{\max}-\sqrt{R\left(\gamma +
\frac{1}{\lambda}\right)} - \frac{1}{\lambda}}{p_{\max} - p_{\min}},
\end{equation}
then $\bfc$ will be contained in the output list of the decoder.
\end{thm}
\begin{proof}
Let $\bfc$ be the transmitted codeword and  let $\bfy$ be the
received vector.
Substituting (\ref{eq:i}) in Lemma \ref{lem:h}, we get
\begin{equation*}\label{eq:j}
\frac{\sum^n_{i=1}\lfloor \lambda w_{c(i),y(i)} \rfloor}
{\sqrt{\sum^n_{i=1}\sum^q_{j=1}\left(\lfloor \lambda
w_{i,j}\rfloor^2+\lfloor \lambda w_{i,j} \rfloor\right)}} \ge
\sqrt{k-1}.
\end{equation*}
Instead of this condition for successful decoding let us use a more
stringent one obtained by removing the integer parts:
\begin{equation}\label{eq:k}
\frac{\sum^n_{i=1} \left(\lambda w_{c(i),y(i)} - 1\right) }
{\sqrt{\sum^n_{i=1}\sum^q_{j=1}\left((\lambda w_{i,j})^2+ \lambda
w_{i,j}\right)}} \ge \sqrt{k-1}.
\end{equation}
 Rearranging and using the channel
statistics $\gamma$ yields
\begin{equation}\label{eq:l}
\frac{1}{n}\sum^n_{i=1}w_{c(i),y(i)} \ge \sqrt{R\left(\gamma +
\frac{1}{\lambda}\right)-\frac{\gamma}{n}-\frac{1}{\lambda n}} +
\frac{1}{\lambda}.
\end{equation}
Inequality (\ref{eq:l}) certainly holds true for the codeword $\bfc$
if
\begin{equation}\label{eq:m}
\frac{1}{n}\sum^n_{i=1}w_{c(i),y(i)} \ge \sqrt{R\left(\gamma +
\frac{1}{\lambda}\right)} + \frac{1}{\lambda}.
\end{equation}
We have derived a condition based on a specific $\bfy$, but we are
interested in ASD's performance for any $\bfy$ when $\bfc$ is
transmitted. Thus, $\bfy$ becomes a random variable. Let $W_i$ be a
random transition probability $w_{c(i),y(i)}$ given random $\bfy$.
Note that the $W_i$s do not depend on $\bfc$ because of the additive
noise assumption. The p.m.f. of $W_i$ is as follows:
\begin{equation*} p_{W_i}(p_i):= \Pr\{{W_i}
= p_i\} = p_i, \quad i=1,2,\dots, q.
\end{equation*}
We can now rewrite (\ref{eq:m}) as follows:
  \begin{align*}
\frac{1}{n}\sum^n_{i=1}{W}_i &\ge \sqrt{R\left(\gamma +
\frac{1}{\lambda}\right)} + \frac{1}{\lambda}.
   \end{align*}
Since the received vector $\bfy$ differs from $\bfc$ in $t$
coordinates, we can bound the left-hand side of the last inequality
below as follows:
  $$
 \sum^n_{i=1}{W}_i\ge
tp_{\min} + (n-t)p_{\max}.
  $$
Indeed, if $y_i\ne c_i$ then the probability $w_{c(i),y(i)}\ge
p_{\min}.$ On the other hand, if there is no error in coordinate $i$
then the probability $w_{c(i),y(i)}=p_{\max}$; otherwise,
communication over the channel would be impossible. Thus, $\bfc$ is
on the soft-decision decoder's list if
\begin{equation*}
\frac{1}{n} (tp_{\min} + (n-t)p_{\max}) \ge \sqrt{R\left(\gamma +
\frac{1}{\lambda}\right)} + \frac{1}{\lambda}.
\end{equation*}
The theorem follows.
\end{proof}

\begin{figure*}\label{fig:b}
\centering \graphicspath{../..}
\includegraphics [width=9 cm] {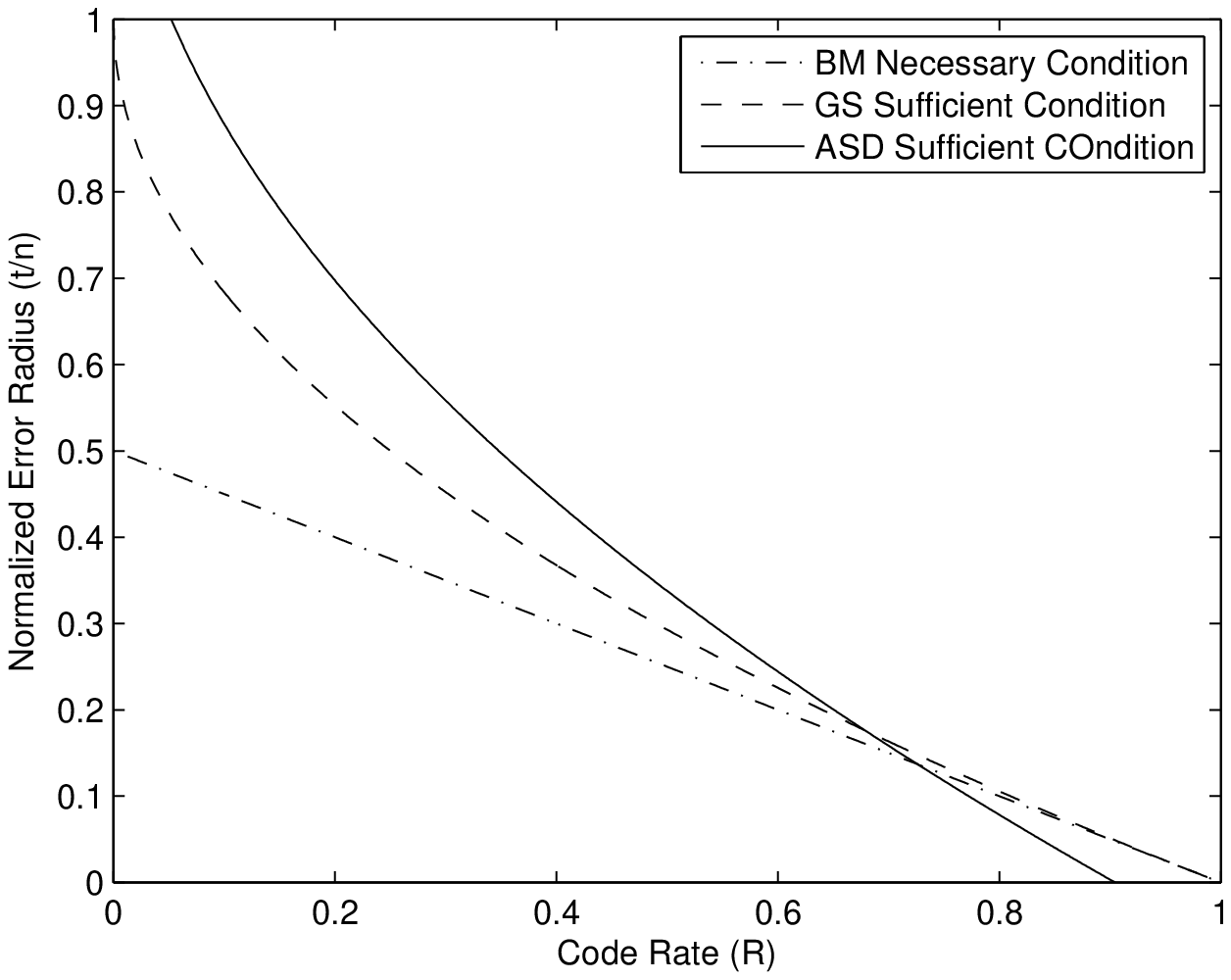}
\end{figure*}
\begin{figure*}
\centering{\mbox{(a)}}
\end{figure*}
\begin{figure*}\label{fig:c}\graphicspath{../..}
\centerline{ \mbox{\includegraphics [width=9 cm] {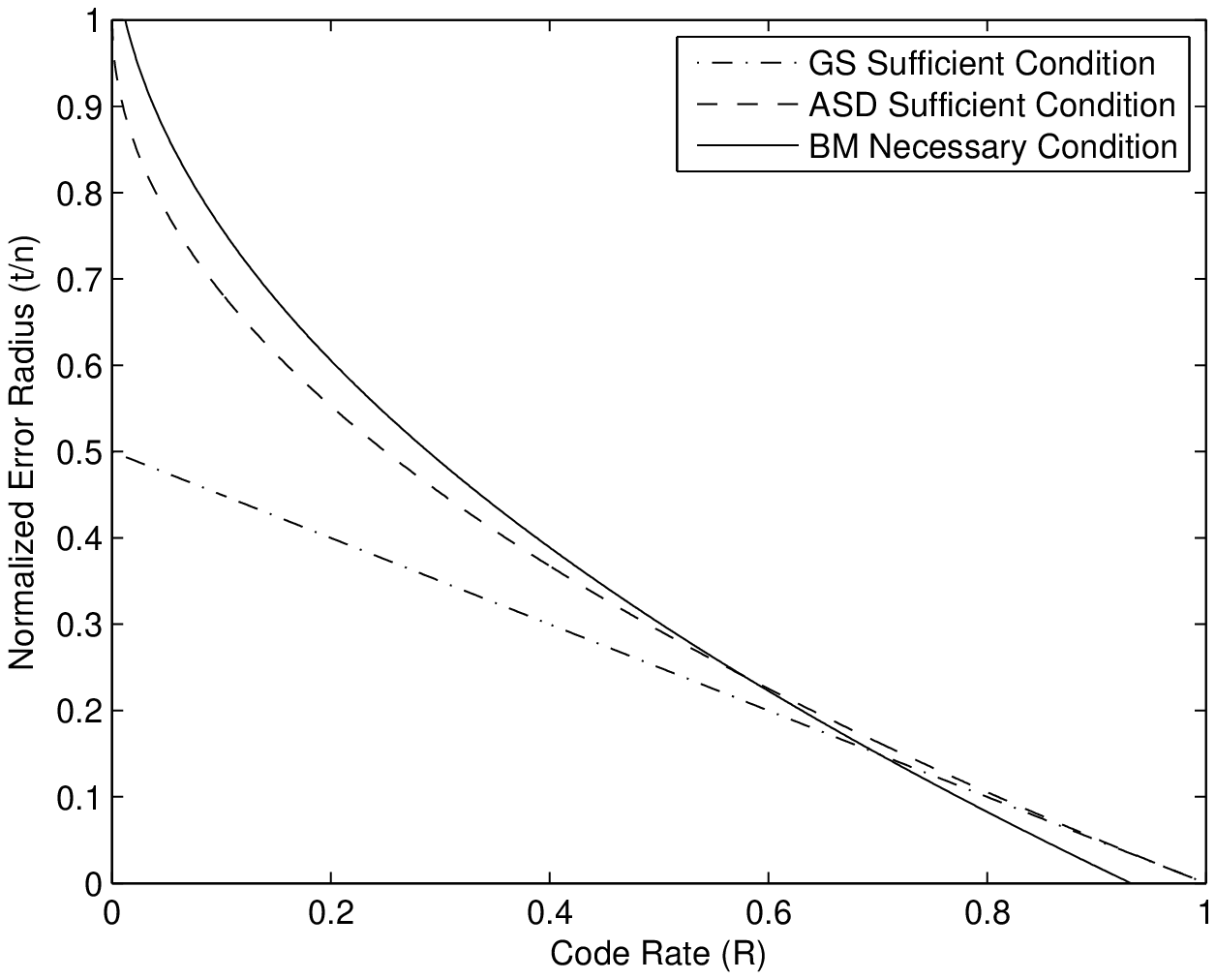}}
\mbox{\includegraphics [width=9 cm] {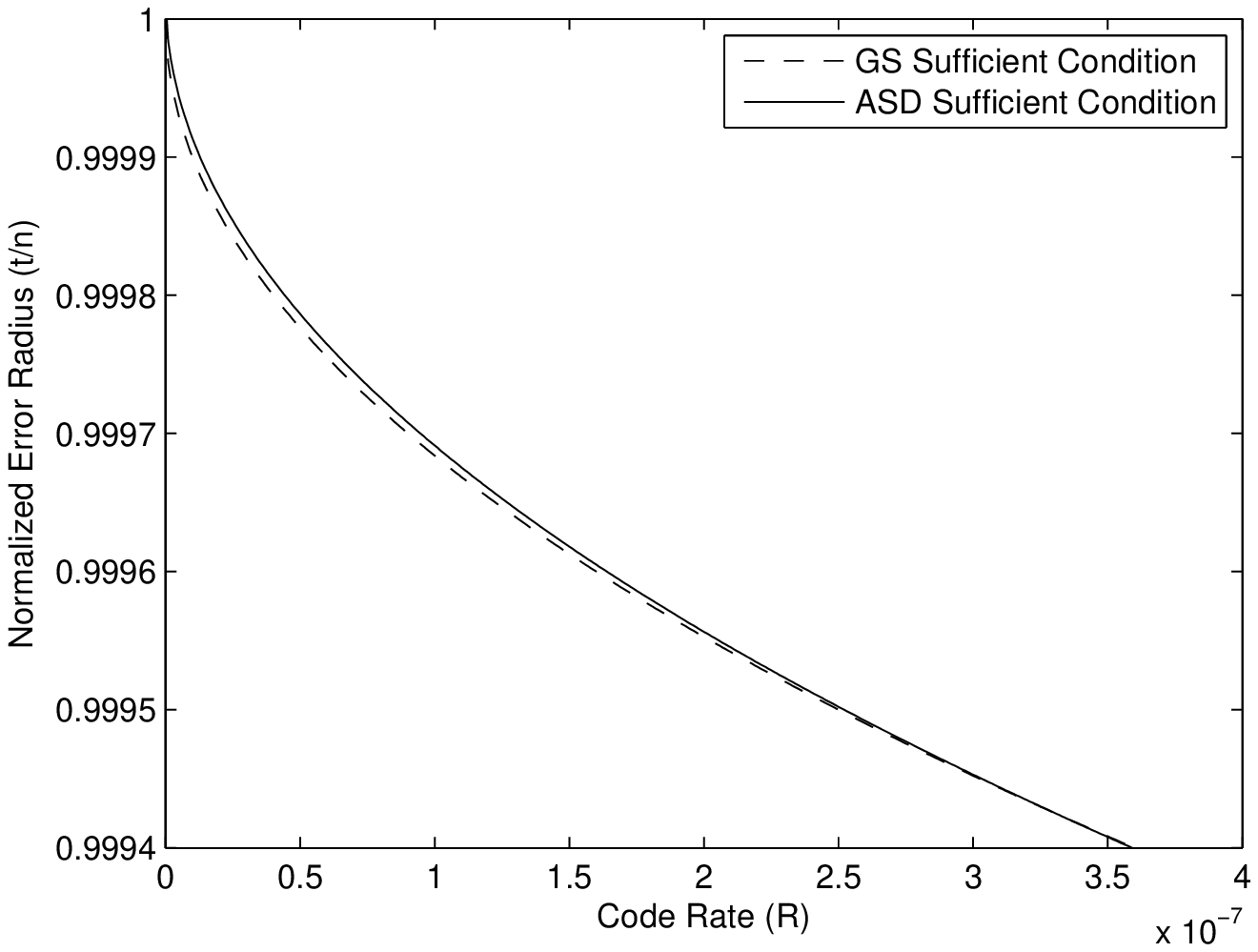}}}
\centering{\mbox{(b) \; \; \; \; \; \; \; \; \; \; \; \; \; \; \; \;
\; \; \; \; \; \; \; \; \; \; \; \; \; \; \; \; \; \; (c)}}
\caption{Decoding radius of ASD compared to GS and BM decoding for
 a), the ``typewriter channel" of Example 1, b), ``two-error channel"
of Example 2, and c), the q-ary symmetric channel of Example 3.}
\end{figure*}

A first look at the error radius in Theorem \ref{thm:a} reveals that
(\ref{eq:t}) becomes large as $p_{\min}$ approaches $p_{\max}$. In
other words, ASD performs well when the channel is far from $q$-ary
symmetric. If the channel is noiseless and $\lambda$ is sufficiently
large, then the bound in Theorem \ref{thm:a} reduces to $1 -
\sqrt{R}$ which is the GS normalized error bound. Let us consider a
 few examples.

\begin{figure*}
\centering \graphicspath{../..}
\includegraphics [width=9 cm] {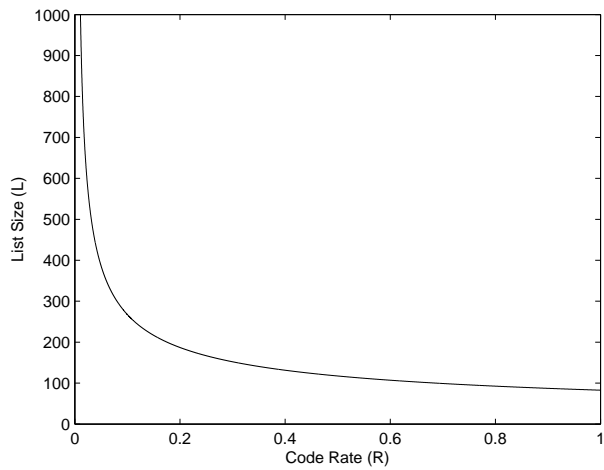}
\caption{List size using ASD for the ``typewriter
channel".}\label{workk}
\end{figure*}

\begin{expl} \label{ex:a}
{\rm Consider the ``typewriter channel" where $w_{i,i} = 0.8$,
$w_{i,j} = 0.2$ for some $j \ne i$, and $w_{i,j} = 0$ for all the
remaining pairs $(i,j)$. Thus, $p_{\max} = 0.8$, $p_{\min} = 0.2$,
and $\gamma = 0.68$. Let us set $\lambda = 100$. Figure 1(a) shows
the normalized error correction radius compared to the GS error
bound for this example. The BM error bound is also shown for
reference. ASD is able to produce a list with the codeword $\bfc$
for a greater error radius than GS decoding for many low to medium
rates. The range of rates for which ASD decoding corrects more
errors than GS decoding is characterized below in this section.}
\end{expl}

\begin{expl}
{\rm Next, let us look at the error radius of Theorem \ref{thm:a}
when there are two possible, equiprobable errors per symbol
transmission, termed the ``two-error channel." In this case,
$p_{\max} = 0.8$, $p_{\min} = 0.1$, $\gamma = 0.66$, and we again
set $\lambda = 100$. Figure 1(b) shows the ASD, GS, and BM curves
for this example.}
\end{expl}

\begin{expl} \label{ex:b}
{\rm Figure 1(c) shows the normalized error bound compared to the GS
error bound for a $q$-ary symmetric channel with $p_{\max} = 0.805$
and $q = 16$. Thus, $p_{\min} = 0.013$ and $\gamma = 0.6506$, and we
set $\lambda = 100$. ASD still provides an improvement, but it is
only for extremely low-rate codes. The lack of improvement for the
$q$-ary symmetric channel is expected since all error patterns,
given that there are $t$ errors, are equally probable.}
\end{expl}

As the channel approaches $q$-ary symmetric, ASD is shown to have a
progressively smaller improvement over GS decoding for low-rate
codes. The fact that the bound on the error radius for ASD is below
the GS radius for higher rates is due to to the approximations used
in the proof of Theorem \ref{thm:a}. The true ASD error radius is
likely to be slightly greater or within a neighborhood of the GS
decoding radius for all code rates.

In \cite{KO}, Koetter considers a general transmission scenario when
an RS or related code is used for communication over a symmetric
channel, and the decoder's performance is measured in terms of an
arbitrary additive cost function $\Delta:\ff_q\to \mathbb{R}^+.$
\cite{KO} attempts at optimizing the assignment of multiplicities
for the purpose of finding the maximum attainable decoding radius.
An error vector $\bfe\in \ff_q^n$ is assigned the cost
   $$
    \Delta(\bfe)=\sum_{i=1}^n \Delta(e_i).
  $$
The result of \cite{KO} stated for an additive channel, is as
follows (the original text contains some misprints).
 \begin{thm}  Suppose that a code vector $\bfc$ of an RS code
of rate $R$ transmitted over the channel is received as
$\bfy=\bfc+\bfe.$ If the values $R$ and $\Delta(\bfe)$
simultaneously satisfy the inequalities
   \begin{align}
   R&\le (1-\epsilon) \sum_{a\in \ff_q} [\rho-\theta \Delta(a)]_+^2
  \quad(\epsilon>0)\label{eq:ko1}\\
    \Delta(\bfe)&\le \sum_{a\in \ff_q} \Delta(a) [\rho-\theta
\Delta(a)]_+
 \label{eq:ko2}
   \end{align}
where $[x]_+=\max(0,x),$ $\theta>0$ is an arbitrary parameter, and
    $$
     \sum_{a\in \ff_q}[\rho-\theta \Delta(a)]_+ = 1,
    $$
then the vector $\bfc$ will be placed on the ASD output list.
Moreover, relations {\rm(\ref{eq:ko1})-(\ref{eq:ko2})} characterize
the optimum tradeoff between the code rate and the cost of error.
\end{thm}

We note that even though the theorem's result is optimum for the
``error radius''  in terms of the cost function, its result is
inferior to the result of Theorem \ref{thm:a} in the following
sense. For the previous theorem to give a performance curve over
Hamming distance, the decoder is optimized over Hamming distance, a
concept that is congruent with a hard-decision decoder, not a
soft-decision decoder. Namely, taking the distance as the cost
function, we have $\Delta(\alpha_i)=1-\delta_{\alpha_i,0},
i=1,\dots, q$ (the cost is 1 for all the symbols of $\ff_q$ except
for some distinguished symbol whose value is of no importance).
Then (\ref{eq:ko1})-(\ref{eq:ko2}) reduce to the equation
  \begin{align*}
       \frac tn&\le \frac n{n+1}-\sqrt{\frac{nR}{(n+1)(1-\epsilon)}
-\frac n{(n+1)^2}}, \\
\frac {1-\epsilon}{n+1} &\le R \le 1 - \epsilon
  \end{align*}
which is an error radius that slightly exceeds the GS bound
(\ref{eq:b}) for most rates. As expected, it is inferior to
(\ref{eq:t}) for a subset of low to medium rate codes.

\subsection{Size of the List}

\begin{prop}\label{thm:b} Assume that $k\ge 2.$
Given an additive noise channel, the size of the list for an
algebraic soft-decision decoder is bounded above by
\begin{equation}\label{eq:pop}
\left\lfloor\sqrt{\frac{n(\lambda^2\gamma + \lambda)}{k-1} +
\left(\frac{k+1}{2k-2}\right)^2}-
\left(\frac{k+1}{k-2}\right)\right\rfloor.
\end{equation}
\end{prop}
\medskip
\begin{proof}
Assuming PMAS (\ref{eq:i}), observe that
\begin{equation*}\label{eq:p}
2\mathcal{C}(\mathcal{M}) = n\sum_{i=1}^q\left\lfloor\lambda
p_i\right\rfloor^2 + n\sum_{i=1}^q\left\lfloor\lambda
p_i\right\rfloor
 \le n \lambda^2 \gamma + n \lambda.
\end{equation*}
Substituting this upper bound in Lemma \ref{lem:him} gives the
result.
\end{proof}

The bound in Proposition \ref{thm:b} is based on the $Y$-degree of
the polynomial $Q(X,Y)$. Thus, it includes polynomials that are a
good fit for the set of multiplicity points, including higher-degree
polynomials. As a result, the bound of Proposition \ref{thm:b} is
not tight.

\begin{expl}
{\rm Figure 2 shows a graph of the bound on the list size presented
in Proposition \ref{thm:b} for Example \ref{ex:a} with $n = 255$. }
\end{expl}

Since the bound in (\ref{eq:pop}) is a strictly decreasing function
of the code's dimension $k$, an upper bound on the size of the list
is obtained by taking $k=2:$
\begin{equation*}\label{eq:pqp}
| \mathcal{L} | \le \left\lfloor\sqrt{n\lambda^2\gamma + n\lambda +
\frac{9}{4}} - \frac{3}{2}\right\rfloor <
\sqrt{n(\lambda^2\gamma+\lambda)}.
\end{equation*}
Thus, the number of codewords on the list is bounded above by a
slowly growing function of $n$.

\subsection{A Closer Look at the ASD Error Radius}

We are interested in quantifying when the error radius of ASD
exceeds the error radius of GS decoding. For simplicity we only
analyze the case of GS decoding with $m\to\infty$ (note that it will
also imply that ASD is better than GS decoding for any finite $m$).

\begin{cor}\label{cor:b}
With $\lambda > \nicefrac{1}{p_{\min}}$, the algebraic soft-decoding
radius exceeds the GS decoding radius if
\begin{equation}\label{eq:r}
R < \Big(\frac{p_{\min}-{1}/{\lambda}}{\sqrt{\gamma+{1}/{\lambda}}-
p_{\max}+p_{\min}}\Big)^2.
\end{equation}
\end{cor}

\begin{proof}
Solving the equation
\begin{equation*}
\frac{p_{\max}-\sqrt{R\left(\gamma + {1}/{\lambda}\right)} -
{1}/{\lambda}}{p_{\max} - p_{\min}} > 1 - \sqrt{R}
\end{equation*}
for $R$ and assuming $\lambda > \nicefrac{1}{p_{\min}}$ yields the
corollary.
\end{proof}

 Corollary \ref{cor:b} gives a sufficient condition for the ASD correction
radius to exceed the GS decoding radius. This condition is nontrivial for a subset of low code rates and $\lambda$
large enough (see Example \ref{ex:a}). One
also notices in Example \ref{ex:a} that there is another non-zero
subset of code rates where the transmitted codeword is always on the
list, i.e. $\nicefrac{t}{n} \le 1$. Corollary \ref{cor:c} quantifies
that region.

\begin{cor}\label{cor:c}
Let $\lambda > \nicefrac{1}{p_{\min}}$. If
\begin{equation*}\label{eq:s}
R \le \frac{\left(p_{\min} - \frac{1}{\lambda}
\right)^2}{\gamma+\frac{1}{\lambda}},
\end{equation*}
then an algebraic soft-decision decoder will always produce a list
containing the transmitted codeword $\bfc$.
\end{cor}

\begin{proof}
If the right-hand side of (\ref{eq:t}) is $1$, then ASD will produce
a list that contains $\bfc$ regardless of the error pattern. Thus,
the claim reduces to solving for $R$ the inequality
\begin{equation*}
\frac{p_{\max}-\sqrt{R\left(\gamma + \frac{1}{\lambda}\right)} -
\frac{1}{\lambda}}{p_{\max} - p_{\min}} \ge 1,
\end{equation*}

\noindent from which the corollary follows.
\end{proof}

It is a surprising result that there exists non-zero rates where ASD
always produces a list polynomial in $n$ that contains the
transmitted codeword. The intuition is that $Pr(\bfy | \bfc) \ne 0$
for all channels, but for many codewords $\bfc' \in C$, $Pr(\bfy |
\bfc') = 0$ due to zeros in the transition probability matrix. The
zeros in the transition probability matrix allow the soft-decision
decoder to discount codewords. When the code rate is low enough, the
decoder can reduce the size of the list to be within the bound of
Proposition \ref{thm:b} without actually eliminating possible
codewords. In this case, the probability of list-decoding error is
zero.

For the $q$-ary symmetric channel, a case where intuition tells us
that ASD should provide no improvement over GS decoding, Figure~1(c)
still shows a region that where ASD's error radius exceeds GS
decoding's error radius.  However, there is no code construction
that simultaneously satisfies the condition $\lambda >
{1}/{p_{\min}}$ and (\ref{eq:r}) for the $q$-ary symmetric channel
unless we have $\lambda \to \infty$. Thus, there is no achievable
rate region where the ASD radius is larger than the GS decoding
radius except when the size of the list is unbounded.

\section{ASD Probability of Error Bounds}\label{sect:pe}

This section is focused on bounding the probability of error for
ASD. In the list-decoding setting, the probability of error has
generally been defined as the probability that the transmitted
codeword is not on the decoder's list. In the final stage of
decoding, it may be required to select a unique codeword candidate
from the list obtained using the maximum likelihood criterion. In
this section, we derive a bound for the probability of error when
the correct decoding corresponds to the case that the transmitted
codeword is on the decoder's list \textit{and} it is selected as the
best estimate.

The only previous work that deals with deriving bounds on the
probability of error for ASD is \cite{RK}. In that paper, Ratnakar
and Koetter consider a general channel and use list-decoding error
as the error event. We refine the results of \cite{RK} in two ways:
first, we prove a tighter bound on the list-decoding error
probability for low rates, and secondly, we derive a bound on the
probability of selection error that is the dominating error event
for those rates.
%

\subsection{General Form}

As in the previous section, the channel is assumed to be additive
and memoryless. The transmitted codeword is $\bfc$, the decoder's
list is $\mathcal{L}$, and the decoder's chosen codeword is $\hat
\bfc$. Define the following random events $\mathcal{A}$ and
$\mathcal{B}$ as
\begin{equation*}
\mathcal{A}: \bfc \notin \mathcal{L}, \qquad \mathcal{B}: \hat\bfc
\not= \bfc.
\end{equation*}

\noindent The list-decoding probability of error, which is the
probability that the list produced by ASD contains the transmitted
codeword, is given by $P\{\mathcal{A}\}$. The selection probability
of error is given by $P\{\mathcal{B}\}$. Observe that $\mathcal{A}
\subseteq \mathcal{B}$, giving us $P\{\mathcal{A}\} \le
P\{\mathcal{B}\}$. Each of these probabilities can be bounded above
by using the Chernoff bound (see e.g. \cite{RG}).

\begin{lem}\label{lem:aa}
($\mathit{Chernoff} \mathit{bound}$) Let $w$ be a random variable
with moment generating function $\Phi_w(s)$ and let A be a real
number. Then
\begin{equation}\label{eq:ee}
\Pr\{w \ge A \} \le e^{-sA}\Phi_w(s), \quad s > 0
\end{equation}
\begin{equation}\label{eq:ff}
\Pr\{w \le A \} \le e^{sA}\Phi_w(-s), \quad s > 0.
\end{equation}
\end{lem}

In applications, one optimizes on the choice of $s$ to obtain the
tightest bound possible. The Chernoff bound will allow us to write
\begin{equation*}
P\{\mathcal{A}\} \le e^{-nE_{\mathcal{A}}}, \quad P\{\mathcal{B}\}
\le e^{-nE_{\mathcal{B}}}.
\end{equation*}
The functions $E_{\mathcal{A}}$ and $E_{\mathcal{B}}$ are the error
exponents for $P \{\mathcal{A}\}$ and $P \{\mathcal{B}\}$,
respectively.

\subsection{List-Decoding Probability of Error}

The next theorem gives an upper bound on the probability that the
transmitted codeword is not on the list output by the decoder. A
similar estimate of the error exponent, in a more general context,
is derived in \cite{RK}. However \cite{RK} does not contain the
analysis of the error radius that leads us to conclude that for low
rates the transmitted codeword will always be on the list. In other
words, the error event for these (and some other) rates will be
dominated by an incorrect selection from the list obtained through
ASD.

\begin{thm}\label{thm:aa}
The probability of event $\mathcal{A}$ can be bounded above as
\begin{equation*}
P \{\mathcal{A}\} \le e^{-nE_{\mathcal{A}}},
\end{equation*}
\noindent where
\begin{equation*}
E_{\mathcal{A}} = -\ln\sum^q_{i=1}p_ie^{-s\left(p_i-\sqrt{R(\gamma +
{1}/{\lambda})} - {1}/{\lambda}\right)}
\end{equation*}
\noindent except when $R < \frac{(p_{\min}-{1}/{\lambda})^2}{\gamma
+{1}/{\lambda}}$ and $\lambda > \nicefrac{1}{p_{\min}}$, in which
case $E_{\mathcal{A}} = \infty$.
\end{thm}

\begin{proof}
The rate region where $E_{\mathcal{A}} = \infty$ follows directly
from Corollary \ref{cor:b}. In order to gain insight into the event
$\mathcal{A}$ for the remainder of the rates, consider again the
threshold condition for ASD list-decoding success given in Lemma
\ref{lem:h}. If the condition in Lemma \ref{lem:h} is met, it is
clear that $\mathcal{A}$ is false. However, if the condition is not
met, $\mathcal{A}$ could either be true or false. Thus, for a given
$\mathcal{M}$, we have

\begin{equation*}\label{eq:unc}
P \{ \mathcal{A} \} \le \Pr \left\{ S_{\mathcal{M}}(\bfc) <
\sqrt{2(k-1)\mathcal{C}(\mathcal{M})} \right\}.
\end{equation*}
Assume now that the received vector $\bfy$ is a random vector with
coordinates distributed according to the transition probabilities in
the channel. As above, let $W_i$ be a random transition probability
$w_{c(i),y(i)}.$ With these assumptions, the components of the
matrix of multiplicities as well as the score of a codeword $\bfc$
and the cost of $\cM$ become random variables. Then
\begin{equation*}
\begin{split}
\Pr \big\{S_{\mathcal{M}}(\bfc) <
&\sqrt{2(k-1)\mathcal{C}(\mathcal{M})} \big\}
 \le \\ & \Pr \Big\{\sum^n_{i=1}{\mathbf W}_i < n\sqrt{R\Big(\gamma +
\frac{1}{\lambda}\Big)} + \frac{n}{\lambda}\Big\}.
\end{split}
\end{equation*}
Finally, the Chernoff bound (\ref{eq:ff}) can be applied to give the
result
\begin{equation*}
P \{ \mathcal{A} \} \le e^{sn\big(\sqrt{R(\gamma + \frac1\lambda)} +
\frac{1}{\lambda}\big)} \Big(\sum^q_{i=1}p_i e^{-sp_i}\Big)^n
\end{equation*}
where $s > 0$. The theorem follows.
\end{proof}

In order to obtain the tightest bound, we need to maximize
$E_{\mathcal{A}}$ through proper choice of $s$. If we define $g(s)$
by $E_{\mathcal{A}} = -\ln(g(s))$, then the goal is to minimize
$g(s)$. When the maximum value of $E_{\mathcal{A}}$ is negative,
then no conclusion can be drawn regarding the possibility of
reliable communication. The following lemma shows that reliable
communication is possible when the code rate is less than a rate
maximum that can be well-estimated by $\gamma$.

\begin{lem}
\begin{align*}
E_{\mathcal{A}} &> 0  \quad\text{if }R < \frac{(\gamma -
{1}/{\lambda})^2}{\gamma + {1}/{\lambda}}
\end{align*}
\end{lem}
\begin{proof}
We have that
\begin{equation*}
g(s) = \sum^q_{i=1}p_ie^{-s\left(p_i-\sqrt{R(\gamma +
\frac{1}{\lambda})} - \frac{1}{\lambda}\right)}.
\end{equation*}

\noindent First observe that $g''(s) > 0$, indicating that the
function $g(s)$ is convex. Next observe that $g(0) = 1$ and that
$g'(0) < 0$ if and only if
 $R < (\gamma -1/\lambda)^2/(\gamma + {1}/{\lambda})$.
In the case that $g'(0) < 0$, the minimum value of $g(s)$ is
achieved for some $s'
> 0$, and since $g(0) = 1$, $g(s') \le 1$. Thus, $E_{\mathcal{A}} >
0$. Otherwise, $g(s) \ge 1$ which gives a trivial estimate
$E_{\mathcal{A}} = 0$.
\end{proof}

\subsection{Selection Probability of Error}

\begin{figure*}\label{fig:aa}
\centering \graphicspath{../..}
\includegraphics [width=12 cm] {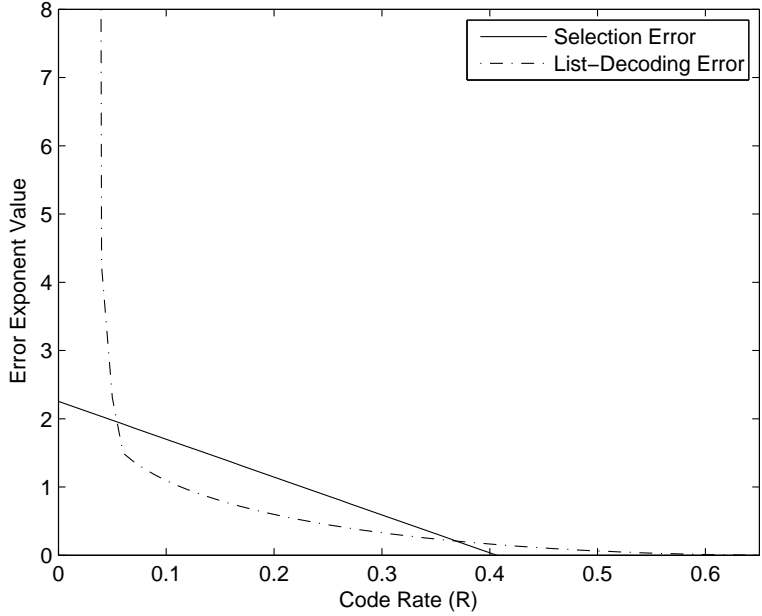}
\caption{Comparison of the two error exponents for ASD over low
rates.}
\end{figure*}

When the list-decoding probability of error is zero as for the rate
region defined in Corollary \ref{cor:c}, one does not have insight
into the performance of ASD. Therefore, it is of interest to
quantify a comprehensive probability of error given by $P \{
\mathcal{B} \}.$ To make the selection error probability amenable to
analysis we consider the ensemble of random Generalized Reed-Solomon
(GRS) codes.

Let $C$ be an RS code and $\bfw=(w_1,\dots,w_n)$ be a vector of
nonzero elements of $\ff_q.$ A GRS$_k(\bfw)$ code is obtained from
$C$ be multiplying every code vector $\bfc\in C$ by a diagonal
matrix $\text{diag\,}(w_1,\dots,w_n).$ Thus, the code $C$ gives rise
to the ensemble of $(q-1)^n$ GRS$_k$ codes with uniform distribution
on it induced by the choice of the modifying vector $\bfw.$

\remove{For a polynomial $f\in \mathbb{F}[x]$ and a vector $\bfw =
(w_1,\dots,w_n) \in \mathbb{F}_q^n$, define the evaluation mapping
$\bfw \textnormal{-eval}:f \to \mathbb{F}^n_q$ given by
$(\bfw\textnormal{-eval} f)_i = w_{i-1} f(\alpha_i)$, $2 \le i \le
q$. The evaluation mapping associates a $q$-ary $n$-vector to every
polynomial $f\in \mathbb{F}_q[x].$

\begin{defn}
Let $\bfal = (\alpha_2,\dots,\alpha_q)$. A $q$-ary Generalized
Reed-Solomon code of length $n=q-1$ and dimension $k$, denoted
$GRS_k(\bfal,\bfw)$, is the set of codewords of the form
  $$
     \{\bfc=\bfw \textnormal{-eval} (f): \;f\in \mathbb{F}_q[x], 0\le
\deg f\le k-1
        \}.
  $$
\end{defn}
We will term a random GRS Code as one where $\bfw$ is chosen
uniformally over $\mathbb{F}_q^n$ before each transmission.}

We will assume transmission with a random GRS code where $\bfw$ is
chosen uniformly before each transmission.
\begin{thm}\label{thm:f}
Suppose that a random GRS code $G$ of rate $R$ is used to transmit
through an additive channel. The decoding error probability of ASD
can be bounded above as
\begin{equation*}
P \{\mathcal{B}\} \le e^{-nE_{\mathcal{B}}}
\end{equation*}
where
\begin{equation*}
E_{\mathcal{B}}= - \ln \Big[
q^{R-1}\big(e^{s/\lambda}+2\sum^q_{i=1}\sum^q_{\substack{j=1
\\ j \ne i}}p_ie^{-s(p_i-p_j-\frac{1}{\lambda})}\big)\Big].
\end{equation*}
\end{thm}

\begin{proof}
Let $\{\bfc_1,\dots,\bfc_M\}$ be the codewords of the code $G$.
Define the events $\mathcal{C}_i$ and $\mathcal{D}_i$ as follows:
  $$
\mathcal{C}_i=\{\left(\bfc_i  \in \mathcal{L}\right) \&
\left(S_\cM(\bfc_i) \ge S_\cM(\bfc)\right)\}
$$
$$
\mathcal{D}_i=\{S_\cM(\bfc_i) \ge S_\cM(\bfc)\}
$$
where $\cL$ is the decoder's list of codewords, $\bfc$ is the
transmitted codeword, and $\cM$ is the multiplicity matrix.

Observe that
\begin{equation*}
\begin{split}\label{eq:cc}
P \{ \mathcal{B} \} & = \Pr \{\exists \bfc' \ne
\bfc, \; \bfc' \in \mathcal{L}: S_\cM(\bfc') \ge S_\cM(\bfc) \} \\
& = P \Big\{ \bigcup^{q^k}_{\substack{i = 1 \\ i:\bfc_i \ne \bfc}}
\mathcal{C}_i \Big\} \le \sum^{q^k}_{\substack{i = 1
\\ i:\bfc_i \ne
\bfc}} P \{ \mathcal{C}_i \} \nonumber\\
& \le \sum^{q^k}_{\substack{i = 1 \\ i:\bfc_i \ne \bfc}} P \{
\mathcal{D}_i \}.
\end{split}
\end{equation*}

For all $\bfv_i \in \mathbb{F}_q^n$, define the event
$\mathcal{E}_i$ as follows:
\begin{equation*}
\mathcal{E}_i:\{S_\cM(\bfv_i) \ge S_\cM(\bfc)\}.
\end{equation*}
We have
\begin{align*}
\sum^{q^k}_{\substack{i = 1 \\ i:\bfc_i \ne \bfc}}
P \{ \mathcal{D}_i \} & = \sum^{q^n}_{\substack{i = 1 \\
i:\bfv_i \ne \bfc}}P\{\mathcal{E}_i , \bfv_i \in G\}.
\end{align*}
Further, every coordinate of $\bfc$ is distributed uniformly in
$\bfF_q$ and therefore, the multiplicity $m_{v(i),i}$ is a uniform
random variable taking values in $\{p_1,\dots,p_q\}.$ Moreover,
$\Pr(\bfv_i\in G)=q^{k-n}.$

We then know that $S_\cM(\bfv)$ is a sum of i.i.d. random variables
$m_{v(i),i}$. As a result, $P\{\mathcal{E}_i\}$ is the same for all
$\bfv_i \ne \bfc$, and the events $\mathcal{E}_i$ and $\{\bfv_i \in
C\}$ are independent, i.e.
$$
\sum^{q^n}_{\substack{i = 1 \\
i:\bfv_i \ne \bfc}}P\{\mathcal{E}_i , \bfv_i \in C\} =
\sum^{q^n}_{\substack{i = 1 \\
i:\bfv_i \ne \bfc}}P\{\mathcal{E}_i \} P \{\bfv_i \in C\}
$$
\remove{ Let us first determine $P \{\bfv_i \in C\}$. A vector
$\bfv_i = (v_1,\dots,v_n)$ is in the code $C$ if $\bfv$ matches
$$
(w_1 f(\alpha_2), \dots , w_n f(\alpha_q))
$$
for any $f \in \mathbb{F} [x]$. Equivalently, if
$$
(w_1,\dots,w_n) = (v_1 / f(\alpha_2), \dots , v_n / f(\alpha_q)),
$$
for any $f \in \mathbb{F} [x]$, then $\bfv_i$ is in the code $C$. In
other words, a random variable chooses among $q^n$ equally-likely
choices of which $q^k$ result in success. Hence, $P \{\bfv_i \in C\}
= \nicefrac{1}{q^{n-k}}$, and

\begin{align*}
\sum^{q^n}_{\substack{i = 1 \\
i:\bfv_i \ne \bfc}}P\{\mathcal{E}_i \} P \{\bfv_i \in C\} &}
$$
 = \sum^{q^n}_{\substack{i = 1 \\
i:\bfv_i \ne \bfc}}  \frac{1}{q^{n-k}}  P \{\mathcal{E}_i \} \le q^k
P \{\mathcal{E}_i | \bfv_i \ne \bfc \}.
$$
Define the random variable $W_i$ as in the proof of Theorem
\ref{thm:a} and let $V_i,i=1,\dots,q^n$ be i.i.d.  r.v.'s with
  $$P(V_i=p_j)=1/q, 1\le j\le q.
   $$
Then
$$
P\{\cB\}\le q^k P \{\mathcal{E}_i  | \bfv_i \ne \bfc \}
$$
\begin{align*} & = q^k \Pr \Big\{
\sum^n_{i=1} \lfloor \lambda {V}_i \rfloor \ge \sum^n_{i=1}\lfloor
\lambda {W}_i \rfloor \Big\} \\
& \le q^k \Pr \Big \{ \sum^n_{i=1} \big(V_i - { W}_i \big) \ge -
\frac{n}{\lambda} \Big \}.
\end{align*}
Now let us apply the Chernoff bound (\ref{eq:ee}). For any $s > 0$,
\begin{multline*}
\Pr  \Big \{ \sum^n_{i=1}  \left( {V}_i - {W}_i \right) \ge -
\frac{n}{\lambda} \Big \} \\
\le e^{\frac{sn}{\lambda}} \Big(\sum^q_{i=1}\frac{1}{q}
e^{sp_i}\Big)^n \Big(\sum^q_{i=1}p_i e^{-sp_i}\Big)^n.
\end{multline*}
Algebraic manipulations and the appropriate definition of
$E_{\mathcal{B}}$ yield the theorem.
\end{proof}
\remove{
\begin{equation*}\label{eq:ffg}
\begin{split}
\Pr  \Big \{ \sum^n_{i=1}  \left( {V}_i - {W}_i \right) & \ge -
\frac{n}{\lambda} \Big \} \\ & \le e^{\frac{sn}{\lambda}}
\Big(\sum^q_{i=1}\frac{1}{q} e^{sp_i}\Big)^n \Big(\sum^q_{i=1}p_i
e^{-sp_i}\Big)^n.
\end{split}
\end{equation*}}

It is of interest to find out the behavior of $E_{\mathcal{A}}$
compared to $E_{\mathcal{B}}$ across all values of $R$. Let us
compare the two error exponents in an example.

\begin{expl}
{\rm Consider once again Example \ref{ex:a} ($p_{\max} = 0.8$,
$p_{\min} = 0.2$, $\lambda = 100$, $q = 256$, and $\gamma = 0.68)$.
Figure 3 shows the comparison of $E_{\mathcal{A}}$ to
$E_{\mathcal{B}}$. 
ne can see that reliable communication, in the
list-decoding sense, is assured for rates less than $0.67$, and
reliable communication, based on the probability of error of
selection, is guaranteed for rates less than $0.41$.}
\end{expl}

\remove{
\section{Error Bound for the ASD Algorithm}\label{sect:pe}
In this section we use the technique developed for the analysis of
the error radius to estimate from above the error probability for
the ASD algorithm. As in the previous section, the channel is
assumed to be additive and memoryless. The next theorem, whose proof
will be presented elsewhere, provides an upper bound on the
probability of the list-decoding error.
\begin{thm}\label{thm:aa}
The probability of the event $\cA$ that the transmitted codeword
will not be included on the output list of ASD algorithm can be
bounded above as follows:
\begin{equation*}
\Pr \{\mathcal{A}\} \le e^{-nE_{\mathcal{A}}},
\end{equation*}
\noindent where
\begin{equation*}
E_{\mathcal{A}} = -\ln\sum^q_{i=1}p_ie^{-s\left(p_i-\sqrt{R(\gamma +
{1}/{\lambda})} - {1}/{\lambda}\right)} \quad (s>0)
\end{equation*}
\noindent except when $R < \frac{(p_{\min}-{1}/{\lambda})^2}{\gamma
+{1}/{\lambda}}$ and $\lambda > \nicefrac{1}{p_{\min}}$, in which
case $E_{\mathcal{A}} = \infty$.
\end{thm}
A similar estimate of the error exponent, in a more general context,
is derived in \cite{RK}. However \cite{RK} did not contain the
analysis of the error exponent that leads us to conclude that for
low rates the transmitted codeword will always be on the list. In
other words, the error event for these (and some other) rates will
be dominated by the incorrect selection of the most probable
codeword from the list obtained through ASD.}

\section{Multivariate Interpolation}\label{sect:mv}

\begin{figure*}\label{fig:ccc}
\centering \graphicspath{../..}
\includegraphics [width=10 cm] {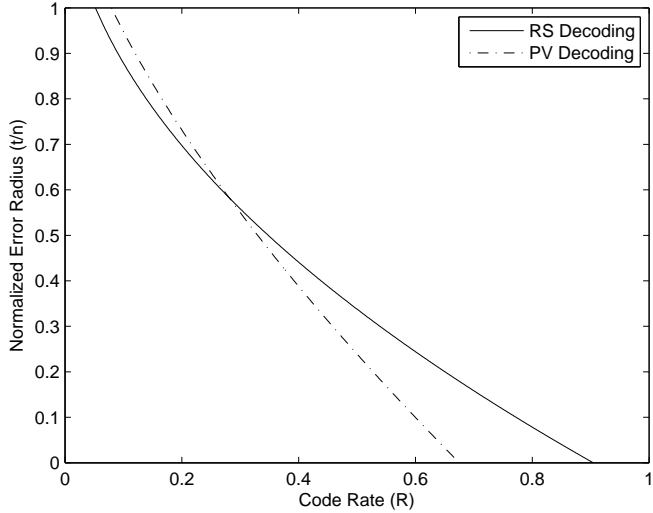}
\caption{Trivariate decoding of a PV code compared to ASD of a RS
code.}
\end{figure*}

In this section, we estimate the decoding radius of ASD for a new
class of codes introduced recently by Parvaresh and Vardy in
\cite{VP}. These codes are constructed as evaluations of $M\ge 2$
polynomials. A multivariate interpolation decoding algorithm for the
codes in \cite{VP} is shown to exceed the GS decoding radius for low
values of the code rate $R$. In this section we extend our analysis
of ASD to multivariate interpolation.

A code $C$ in the Parvaresh-Vardy (PV) family is defined as follows.
Let $\{1, \beta_1, \beta_2, \dots, \beta_{M-1}\}$ be a basis of
$\mathbb{F}_{q^M}$ over $\ff_q$, let $\{a_1, a_2, \dots, a_{M-1}\}$
be a set of positive integers greater than $1$, and let $e(X)$ be an
irreducible polynomial over $\mathbb{F}_q$.
Given a message $\bfu,$ the encoder constructs $f(X)$ as the
polynomial derived from it and finds the set of polynomials
$\{g_1(X), g_2(X), \dots, g_{M-1}(X)\}$ by computing
\begin{equation}\label{eq:b1}
g_i(X) = (f(X))^{a_i} \; \; \textnormal{mod} \; e(X).
\end{equation}

A codeword $\bfc = \{c_1, c_2, \dots, c_n \}$ of a PV code that is
associated with $\bfu$ is found through the evaluation
\begin{equation*}
c_i = f(x_i) + \sum^{M-1}_{j=1}\beta_jg_j(x_i), \;\;\;\;\forall i:1
\le i \le n.
\end{equation*}

It follows that the rate of a PV code $C$ is $R = {k}/{Mn}$ and the
minimum distance is $d = n - k + 1$. Since (\ref{eq:b1}) is a
non-linear operation, the code $C$ is not necessarily linear.

\subsection{Soft-Decision Decoding of PV Codes}

Although Parvaresh and Vardy only considered hard-decision decoding
in \cite{VP}, soft-decision decoding of Folded RS Codes, a broader
class of codes that contains PV codes, was considered by Guruswami
and Rudra in \cite{GR}. Remark 4 of \cite{GR} includes a condition
for list-decoding success. In the next theorem we present a more
stringent condition for list-decoding success of a PV code
transmission.


\begin{thm}\label{thm:yu}
Let $C$ be an $[n,k]$ PV code over $\mathbb{F}_{q^M}$ communicated
over a discrete, memoryless channel with additive noise. Suppose
that it is decoded using a multivariate version of the ASD
algorithm. A codeword $c=(c_1,\dots,c_n)$ will be included in the
list output by the algorithm if
\begin{equation*}
\sum^n_{i=1}m_{c(i),i} > \sqrt[M+1]{(k-1)^M \sum_{i,j}
\binom{m_{i,j}+M}{m_{i,j}-1}}.
\end{equation*}
where $m_{i,j}$ is an element of the $q^M \times n$ multiplicity
matrix $\mathcal{M}$.
\end{thm}

\begin{proof}
By extending equation (38) of \cite{PV}, we can derive an upper
bound for the weighted degree of the multivariate polynomial as
\begin{equation}\label{eq:phg}
\begin{split}
\textnormal{wdeg}\;  Q(X,& Y_1, \dots,Y_M)  \\&< \left\lfloor \sqrt
[M+1] {(k-1)^M \sum_{i,j}\prod^M_{l=0}(m_{i,j}+l)}\right\rfloor.
\end{split}
\end{equation}
where wdeg $X^iY_1^{j_1}...Y_M^{j_M} = i + (k-1)\sum^M_{l=1}j_l$. If
the score $S_{\mathcal{M}}(\bfc)$ exceeds the RHS of (\ref{eq:phg}),
then $\bfc$ is on the algebraic soft-decision decoder's list by an
argument similar to the one employed to prove Lemma \ref{lem:h}.
\end{proof}

\subsection{Multivariate Error Decoding Radius}

Suppose the PV code is transmitted over a channel with transition
probabilities $\{p_1, p_2, \dots, p_{q^M}\}$. The statistics
$p_{\min}$, $p_{\max}$, and $\gamma$ are defined as before over this
new set of transition probabilities. An error radius is given in
Theorem \ref{thm:p} for soft-decision decoding of PV codes.

\begin{thm}\label{thm:p}
Given a PV code with rate $R = \nicefrac{k}{Mn}$ is used to
communicate over an additive-noise channel. If
\begin{equation}\label{eq:tut}
\frac{t}{n} \le \frac{p_{\max}-\sqrt [M+1]
{\frac{R^MM^M}{(M+1)!}\sum^{q^M}_{i=1}\prod^M_{l=0} (p_i +
l/\lambda)}- \frac{1}{\lambda}}{p_{\max} - p_{\min}},
\end{equation}
then an algebraic soft-decision decoder, with complexity factor
$\lambda$, will produce a list that contains the transmitted
codeword $\bfc$.
\end{thm}

The proof is very similar to the proof of Theorem \ref{thm:a} and is
omitted. The multivariate ASD error radius is larger than the
bivariate ASD error radius for low-rate codes. This claim is shown
through an example.

\begin{expl}
{\rm Let us return to the typewriter channel, $p_{\max} = 0.8$,
$p_{\min} = 0.2$, $\gamma = 0.68$, and $\lambda = 100$, and compare
trivariate soft-decision decoding of PV codes to bivariate
soft-decision decoding of RS codes (in other words, ASD). Figure 4
shows the error radii $(\ref{eq:t})$ and $(\ref{eq:tut})$. The graph
shows that trivariate decoding provides an improvement over the
bivariate one for rates less than $0.3$.}
\end{expl}

\section{Conclusion}

The results presented in this paper have shown that soft-decision
algebraic list decoding of RS and related codes and is able to
outperform its hard-decision counterparts for low-rate to
medium-rate codes. An estimate of the error decoding radius derived
in the paper enables ASD to be compared for the first time to other
RS decoding methods. This result has also been extended to
multivariable RS codes. A better estimate of the error probability
for list decoding under ASD is derived which shows that at lower
rates, the list decoding error is not an adequate performance
criterion for this algorithm. A comprehensive probability of error
bound is also derived that includes the previously overlooked
probability of selection error.

An open question that remains unanswered is if ASD's performance
makes it a worthwhile decoder to use in Reed-Solomon coding
applications. For low-rate coding applications with channels that
are far from $q$-ary symmetric, ASD shows the potential to correct a
greater number of errors than hard-decision decoders. However, an
interesting (and important in applications) fact of ASD decoding
outperforming its hard-decision counterparts for high-rate codes
claimed in some experimental studies, so far has not been confirmed
by theoretical analysis.

\renewcommand{\baselinestretch}{0.9}


\small\normalsize
\begin{thebibliography}{15}
\setlength{\parskip}{0.5em}

\bibitem {RB} R. E. Blahut, \emph{Theory and Practice of
Error-Correcting Codes}, Reading, MA: Addison-Wesley, 1983.

\bibitem {CT} T. Cover and J. Thomas, \emph{Elements of
Information Theory}, New York, N.Y.: John Wiley and Sons, 2001.

\bibitem {EM} M. El-Khamy and R. McEliece, ``Interpolation
multiplicity assignment algorithms for algebraic soft-decision
decoding of Reed-Solomon codes," in: A. Ashikhmin and A. Barg
(Eds.), \emph{Algebraic Coding Theory and Information Theory},
pp.~99-120, Providence, RI: AMS, 2005.

\bibitem {RG} R. Gallager,  \emph{Information Theory and Reliable
Communication}, New York, N.Y.: John Wiley and Sons, 1968.

\bibitem {GK} W. Gross, F. Kschischang, R. Koetter, and G. Gulak,
``Applications of algebraic soft-decision decoding of Reed-Solomon
codes," {\em IEEE Trans. Commun.\/}, vol.~54, no.~7, pp.~1224-1234,
2006.


\bibitem {GR} V. Guruswami and A. Rudra, ``Explicit capacity-achieving
list-decodable codes," Electronic Colloquium on Computational
Complexity, Report 05-133, 2005, online at
http://eccc.hpi-web.de/eccc/.

\bibitem {GS} V. Guruswami and M. Sudan, ``Improved decoding of
Reed-Solomon and algebraic-geometric codes," \emph{IEEE Trans.
Inform. Theory\/}, vol.~45, no. 5, pp.~1757-1767, 1999.


\bibitem {JN} J. Jiang and K. Narayanan, ``Performance analysis of
algebraic soft decoding of Reed-Solomon codes over binary symmetric
and erasure channels," \emph{Proc. 2005 IEEE Internat. Sympos.
Inform. Theory\/},  Adelaide, Australia, Sept.~4--9, p. 1186-1190,
2005.


\bibitem {JU} J. Justesen, ``Soft-decision decoding of RS codes,"
\emph{Proc. 2005 IEEE Internat. Sympos. Inform. Theory\/},
Adelaide, Australia, Sept.~4--9, 2005, pp.~1183-1185.

\bibitem {KO} R. Koetter, ``On optimal weight assignments for
multivariate interpolation list-decoding," {\em Proc. 2006 IEEE
Information Theory Workshop\/}, Punta del Este, Uruguay, March
13-17, pp. 37--41, 2006.


\bibitem {KV} R. Koetter and A. Vardy, ``Algebraic soft-decision
decoding of Reed-Solomon codes," {\em IEEE Trans. Inform. Theory\/},
vol.~49, no.~11, pp. 2809-2825, 2003.

\bibitem {MC} R. McEliece, ``The Guruswami-Sudan decoding algorithm
for Reed-Solomon Codes,"  manuscript, 2003, available on-line at
http://www.systems.caltech.edu/EE/Faculty/rjm/.

\bibitem {PV} F. Parvaresh and A. Vardy, ``Multiplicity assignments
for algebraic soft-decision decoding of Reed-Solomon Codes,"
\emph{Proc. 2005 IEEE Internat. Sympos. Inform. Theory\/}, Yokohama,
Japan, June 29--July 3, 2003, p. 250.

\bibitem {VP} F. Parvaresh and A. Vardy, ``Correcting errors beyond the
Guruswami-Sudan radius in polynomial time," {\em Proc. 2005 IEEE
Annual Symposium on the Foundations of Computer Science (FOCS)\/},
pp. 246-257, 2005.

\bibitem {RK} N. Ratnakar and R. Koetter, ``Exponential error bounds
for algebraic soft-decision decoding of Reed-Solomon codes,"
\emph{IEEE Trans. Inform. Theory\/}, vol. 51, no. 11 pp. 3899-3917,
2005.

\end{thebibliography}
\end{document}